# Covering Rough Sets From a Topological Point of View

Nguyen Duc Thuan, Member, IACSIT

*Abstract*—Covering-based rough set theory is an extension to classical rough set. The main purpose of this paper is to study covering rough sets from a topological point of view. The relationship among upper approximations based on topological spaces are explored.

## I. INTRODUCTION

Rough set theory is a mathematical tool to deal with vagueness and uncertainty of imprecise data. The theory introduced by Pawlak (1982) has been developed and found applications in the fields of decision analysis, data analysis, pattern recognition, machine learning, expert systems, and knowledge discovery in databases, among others.

The original rough set theory introduced by Pawlak was based on an equivalent relation on a finite universe U. For pratice use, there have been some extentions on Pawlak's original concept. One extention is to replace the equivalent relation by a arbitrary binary relation and coverings [1,2,3,15]; the other direction is to study rough set via topological method [1-10]. In this work, we construct topology for a family covering rough sets. After that, we study the relationship among upper approximations based on this topological space so that we can study rough set theory by method of topology.

The paper is organized as follows. Section 2 briefly introduces method of topological construction for a family covering rough sets. Section 3, we compare some upper approximations based on this topological space. At last, the paper is concluded with a summarization in section 4.

## II. PRELIMINARIES

In this section, we first recall relevant fundamental concepts. The approximations based on topological space also are presented. Finally, we introduce method of topological contruction for a family covering rough sets.

### A. Covering

Let U be a universe, C a family of subsets of U. C is called a cover of U if no subset in C is empty and $\cup C = U$. The order pair (U, C) is called a covering approximation space if C is cover of U.

Let (U, C) is a covering approximation space, $x \in U$, the minimal description of x is defined as
$Md(x) = \{K \mid x \in K \in C \wedge (\forall S \in C \wedge x \in S \subseteq K \Rightarrow K = S)\}$

### B. Topological space

A topological space is a pair $(U, \tau)$ consisting of a set U and family $\tau$ of subset of U satisfying the following conditions:

($T_1$) $\varnothing \in \tau$ and $U \in \tau$

($T_2$) $\tau$ is closed under arbitrary union.

($T_3$) $\tau$ is closed under finite intersection.

The subsets of U belonging to $\tau$ are called open set in the space, and the complement of the subsets of U belonging to $\tau$ are called closed set; the family $\tau$ of open subsets of U is also called a topology for U.

- A family $\beta \subseteq \tau$ is called a base for $(U, \tau)$ iff every non_empty open subset of U can be represented as a union of subfamily of $\beta$. A family $S \subseteq \tau$ is called a subbase iff the family of all finite intersections is a base for $(U, \tau)$.

- $\overline{A} = \cap \{F \subseteq U : A \subseteq F$ and F is closed$\}$ is called the $\tau$-closure of a subset $A \subseteq U$.

- $A^0 = \cup \{G \subseteq U : G \subseteq A$ and G is open$\}$ is called the $\tau$-interior of a subset $A \subseteq U$.

- $A^b = \overline{A} - A^0$ is called the $\tau$-boundary of a subset $A \subseteq U$.

Let $X \subseteq U$, $\overline{X}$, $X^0$, $X^b$ be closure, interior, and boundary points respectively. X exact if $X^b = \varnothing$, otherwise X is rough.

### C. Binary relation

Suppose R is an arbitrary binary relation on U, the pair (U,R) is called an approximation space. With respect to R, we can define the R-left and R-right neigborhoods of an element x in U as follows:

$l_R(x) = \{y \mid y \in U, yRx\}$ and $r_R(x) = \{y \mid y \in U, xRy\}$

respectively. To construct the topology $\tau_1$ using R-right neigborhoods (similarly, to construct the topology $\tau_2$ using R-left neigborhoods), we consider the family $S_1 = \{r_R(x) \mid x \in U\}$ as a subbase. We can write $S_x = \{G \in S_1 \mid x \in G\}$.

### D. Topological space is constructed from a family of cover

A information system is a pair S = (U,A), where U is a non-empty, finite set of objects and is called the universe and A is a non-empty, finite set of attributes. For each $a \in A$, $a: U \to V_a$, where $V_a$ is the domain of $a$. Let Ra be a general binary as follows:

$x\ Ra\ y$ iff $a(x) \cap a(y) \neq \varnothing$

With this definition, Ra determines a cover Ca of U and a topology $\tau_a$ induced binary relation $R_a$ (*see C*). For all attributes in A, we get the topology $\tau_S$ which induced by subbase $\bigcup_{a \in A} S_a$. Note $S_a$ as a subbase of topology $\tau_a$.

(U, $\tau_S$) is a topological space is constructed from a family of cover $\{C_a \mid \forall a \in A\}$ which induced from $\{R_a \mid \forall a \in A\}$. We call $\tau_S$ is covering which produced from information system S.

### E. *The notion of reducts to the topological space of the binary relation*

In [5] the notion of reducts to the topological space of the binary relation was generalized. Let $P \subseteq A$ be a subset of A, r $\in P$, where A be a class of binary relations right neigborhood of each of whom has noempty finite intersection, r is said to be superfluous binary relation in P if:

$\beta_P = \beta_{(P-\{r\})}$

The set M is called a minimal reduct of P iff:
1. $\beta_M = \beta_P$,
2. $\beta_M \neq \beta_{(P-r)}$, $\forall r \in M$

where $\beta_P$ is base for (U,$\tau$), which is construted by P be a class of binary relations right neigborhood.

### F. *Generalized rough sets induced by coverings*

Let (U,C) be a covering approximation space. $N(x) = \cap\{K \in C \mid x \in K\}$. (U,$\tau$) be a topological space using R-right neigborhoods. We can list some approximation operators in the following Table I:

TABLE I APPROXIMATION OPERATORS

| W.Zhu (1) | |
|---|---|
| $X_+ = \cup\{K \in C \mid K \subseteq X\}$ | $X^+ = X_+ \cup \{N(x) \mid x \in X - X_+\}$ |
| Xu, Zhang (2) | |
| $C_+X = \{x \in U \mid (\cap Md(x)) \subseteq X\}$ | $C^+X = \{x \in U \mid (\cap Md(x)) \cap X \neq \emptyset\}$ |
| Yao (3) | |
| $\underline{X} = \bigcup_{r_R(x) \subseteq X} r_R(x)$ | $\overline{X} = ((\underline{X^C})^C)$ |
| Yao (4) | |
| $\underline{R}X = \{x \in U \mid r_R(x) \subseteq X\}$ | $\overline{R}X = \{x \in U \mid r_R(x) \cap X \neq \emptyset\}$ |
| A.M. Kozae, A.A. Abo Khadra, T. Medhat (5) | |
| $t\underline{X} = X^0$ | $t\overline{X} = \cap\{F \subseteq U : X \subseteq F \wedge F \text{ is closed}\}$ |

## III. THE RELATIONSHIP AMONG UPPER APPROXIMATIONS BASED ON TOPOLOGICAL SPACES

With the genaralized rough sets induced by arbitrary binary relation, the following properties hold

**Proposition 3.1**[1] If R, $R^{-1}$ are two binary relations on U as defined:

x Ry if and only if y$\in$ N(x), y$R^{-1}$x if and only if xRy then
(1) R is reflexive and transitive;
(2) $C^+X = \overline{R}X$ and $C_+X = \underline{R}X$, for all X $\subseteq$ U;
(3) Conversely, for each reflexive and transivitive binary relation R on U, there exists a covering approximation space (U,C) such that $C^+X = \overline{R}X$ and $C_+X = \underline{R}X$, for all X $\subseteq$ U
(4) $X^+ = \overline{R^{-1}}X$, for all X $\subseteq$ U
(5) $C^+X = \bigcup_{x \in X} l_R(X)$ and $X^+ = \bigcup_{x \in X} r_R(x)$ for all X $\subseteq$ U.
(6) if $C^+ = X^+$ for all X $\subseteq$ U, then R is an equivalence relation on U, therefore, $C^+ = X^+$ are Pawlak rough set upper approximations.

Note: $N(x) = \bigcap_{x \in K \in C} K$

**Lemma 3.1** [1] Suppose that C is an unary covering of U, then there exists some reflexive and transitive binary relation R on U such that $Md(x) = \{r_R(x)\}$, where $r_R(x)$ is the R-right neigbohood of an element x in U. Conversely, if R is a reflexive and transitive binary relation on U, then $C = \{r_R(x)\}$ is unary covering of U and $Md(x) = \{r_R(x)\}$ for all x$\in$U.

Now we consider the following two interesting subsets of P(U):

$G = \{X \mid X \in P(U), \overline{R}X = \emptyset\}$

and

$H = \{X \mid X = \overline{R}Y \text{ for some } Y \in P(U)\}$

**Proposition 3.2** If R is transitive then
1. $\overline{R}$ is idempotent i.e $\overline{R}\overline{R} = \overline{R}$
2. $G \cap H = \emptyset$

*Proof*: (1) If R is transitive, for all X$\subseteq$U, we have
$\forall x \in \overline{R}(X) \Leftrightarrow r_R(x) \cap X \neq \emptyset \Leftrightarrow \exists y \in r_R(x) \wedge y \in X \Leftrightarrow$ xRy$\wedge$y$\in$X. So that, $\forall x \in \overline{R}\overline{R}(X) \Leftrightarrow r_R(x) \cap \overline{R}(X) \neq \emptyset \Leftrightarrow \exists y \in \overline{R}X \wedge xRy \Leftrightarrow \exists z \in X \wedge yRz \wedge xRy \Leftrightarrow \exists z \in X, xRz$ (R is transitive) $\Leftrightarrow x \in \overline{R}X$. In other words, $\overline{R}$ is idempotent i.e $\overline{R}\overline{R} = \overline{R}$

(2) Suppose that X$\in$G$\cap$H, then there exist some Y$\in$P(U) such that X= $\overline{R}Y$ and $\overline{R}X = \emptyset$. Thus X=$\overline{R}Y = \overline{R}\overline{R}Y = \overline{R}X = \emptyset$.

In next section, we consider a special covering: $C = \tau_S$, where (U, $\tau_S$) is a topological space is constructed in II.D.

Obviously, proposition 3.1 be still in true. We can explore some new results:

**Proposition 3.3** Let (U,R) be approximation space, R is reflexive binary relation. The Yao (3) approximations and Yao (4) approximations are identical.

*Proof*: Since R is reflexive, we have $\underline{X} = \underline{R}X$, we can write
$\underline{X^C} = \{x \in U \mid r_R(x) \subseteq X^C\} = \{x \in U \mid r_R(x) \cap X = \emptyset\}$, therefore
$(\underline{X^C})^C = \{x \in U \mid r_R(x) \cap X = \emptyset\}^C = \{x \in U \mid r_R(x) \cap X \neq \emptyset\} = \overline{R}X$.

**Proposition 3.4** Let (U, $\tau_S$) be a topological space is constructed in II.D. We consider a special covering of U is C = $\tau_S$. We have $X_+ = t\underline{X}$, for all X$\subseteq$U.

*Proof*: It can easily be seen that all $K \in C$, $C = \tau_S$, so that K is open. With definitions of $X_+$, $t\underline{X}$, we have this proposition.

**Corollary 3.1** Let $(U, \tau_S)$ be a topological space is constructed in II.D. We consider a special covering of U is $C = \tau_S$. Suppose that C is an unary covering of U, then

(1) $X_+ = C_+X = \underline{X} = \underline{R}X = t\underline{X}$

(2) $C^+X = \overline{R}X$

*Proof*: According to lemma 3.1, when C is unary covering, we have C={Md(x) | x∈U} = {$r_R(x)$} x∈U} and |Md(x)| =1, ∀x∈U. Along with proposition 3.1, 3.2, (1) be true. In other words,

$X_+ = C_+X = \underline{X} = \underline{R}X = t\underline{X}$.

We also have $C^+X = \overline{R}X$, because Md(x) = $r_R(x)$.

Like this, we show the relationship of Zhu, Xu and Zhang, Yao(4) approximations. So, the relationship between Yao(3), A.Mkozae et al and other authors is not presented. We can see it in the following properties:

**Proposition 3.5** Let $(U, \tau_S)$ be a topological space is constructed in II.D. We consider a special covering of U is $C = \tau_S$. Suppose that C is an unary covering of U, then

(1) $C^+X \neq \overline{X}$

(2) $\overline{X} = t\overline{X}$

Proof: (1) we can get (1) in the following a counter example:

Suppose that U={a,b,c,d}, topo τ is defined on U, τ = {∅, U, {d},{c,d}}, X= {a,b}.

By definitions of $C^+X, \overline{X}$, we have:

$\overline{X} = ((\underline{X^C})^C) = (\underline{\{c,d\}})^C = \{c,d\}^C = \{a,b\}$

$C^+X = \{a,b,c,d\}$.

Obvious, $C^+X \neq \overline{X}$.

(2) By definition of $\overline{X}$ and C is an unary covering of U, we can write:

$(\underline{X^C})^C = (\bigcup_{Md(x) \subseteq X^C} Md(x))^C = \bigcap_{Md(x) \subseteq X^C} (Md(x))^C$
$= \bigcap_{X \subseteq (Md(x))^C} (Md(x))^C$

Since Md(x)∈ τ is open, Md(x)$^C$ is Closed. In other words,

$(\underline{X^C})^C = \bigcap_{X \subseteq (Md(x))^C} (Md(x))^C = t\overline{X}$.

**Remark 3.1** In general case, two approximations are different. A.M. Kozae, A.A. Abo Khadra, T. Medhat said that their approximations are better than Yao's approximations (3) because their approximations decrease the boundary region by increasing the lower approximation and decreaseing the upper approximation [5]. However, they have not proved it. So, we can see this prolem easily in the following lemma

**Lemma 3.2** Let $(U, \tau_S)$ be a topological space is constructed in II.D. We consider a special covering of U is $C = \tau_S$. For every subset X of U, X⊆ P(U), then

$$t\overline{X} \subseteq \overline{X}$$

*Proof* By the definition of $t\overline{X}$, it is minimal closed set contains X. Also, by the definition $\overline{X}$, it is a closed set contains X. Obvious, $t\overline{X} \subseteq \overline{X}$. From this lemma, we can get

$$t\overline{X} - t\underline{X} \subseteq \overline{X} - \underline{X}.$$

because $t\underline{X} = \underline{X}$.

With respect to coverings are topologies which constructed in II.D, we realize that to reduce coverings are very difficult. Maybe we use idea of Zhu and Wang for this problem [2-3]

In our opinion, before we process approximations of Yao(3) or A.M. Kozae, A.A. Abo Khadra, T. Medhat (5), we should reduce the topological space of the binary relation by II.E

In [1], Guilong Liu, Ying Sai proposed the concept of transformation of coverings and their properties. The transformation of coverings can be seen as a kind of reduction of covering.

Let **C**(U) denote the set of all coverings of U, we define the transformation F from **C**(U) to **C**(U) as follows,

F: **C**(U) → **C**(U), F(C) = C' = {N(x) | x∈U}

This transformation of coverings makes the upper approximations $X^+, C^+X$ and the lower approximation $C_+X$ to be invariant. But, this transformation has not conservation of topological space. In other words, we can not consider the upper approximations and lower approximations of Yao(3) and A.M. Kozae, A.A. Abo Khadra, T. Medhat (5) with this transformations. We can see in the following a counter example,

Suppose that U={a,b,c,d}, topo τ is defined on U,
C = τ = {∅, U, {d}.{c,d}},
F(C)= {N(a)= U, N(b)=U, N(c) = {c,d}, N(d) = {d}}.
Obvious, F(C) be not topology.

## IV. CONCLUSION

In this paper, we study covering rough sets from a topological point of view. The relationship among upper approximations based on topological spaces are explored. In fact, we summarize properties of five general approximation operators. However, approximation operators are discussed on topological space which is constructed from a family of cover induced by an information system. In next time, topological theory of rough set need to be further studied.

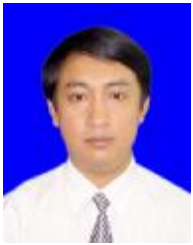

**Nguyen Duc Thuan** was born in Hue, Vietnam, 1962.Brief Biographical History:
-1985, Bachelor in Mathematics- Lecturer of Hue University, Vietnam.
-1998, Master in Information Technology, Lecturer of Nha Trang University, Vietnam.
-2006, Ph.D Student in Institute of Information Technology, Vietnamese academy of Science and Technology.
Current research: Rough Set, Datamining, Distributed Database...